\documentstyle[12pt,epsfig]{article}
\newcommand\slv{v\kern-5pt\raise1pt\hbox{$\scriptstyle/$}\kern1pt}

\def\be{\begin{equation}}
\def\ee{\end{equation}}
\def\bea{\begin{eqnarray}}
\def\eea{\end{eqnarray}}
\newcounter{saveeqn}
\newcounter{App} %\setcounter{App}{0}

\begin{document}
\thispagestyle{empty}
\vspace{0.5cm}
\begin{center}
{\Large \bf $B \to \pi \pi$  Decay in QCD}\\[.3cm]
\vspace{1.7cm}
{\sc \bf Alexander Khodjamirian~$^{a}$ }\\[1cm]
{\em Theory Division, CERN, CH-1211 Geneva 23, Switzerland  \\
and\\
Department of Theoretical Physics, Lund University,\\ 
S\"olvegatan 14A, S - 223 62 Lund, Sweden\\}
\end{center}
\vspace{2cm}

\begin{abstract}
\noindent 

A new method is suggested to calculate the $B\to \pi\pi$ hadronic matrix 
elements from QCD light-cone sum rules.
To leading order in $\alpha_s$ and $1/m_b$, the sum rule reproduces 
a factorizable matrix element, in accordance with the prediction of 
the QCD factorization approach. 
Whereas the QCD factorization can only take into account the 
nonfactorizable corrections induced by 
hard-gluon exchanges, the method suggested here also allows a systematic
inclusion of soft-gluon effects. In this paper, 
I concentrate on the latter aspect and present a calculation
of the nonfactorizable soft-gluon exchange contributions to  $B\to \pi\pi$. 
The result, including twist 3 and 4 terms, is suppressed by one power 
of $1/m_b$ with respect to the factorizable amplitude. 
Despite its numerical smallness, the soft effect is at the same level as 
the $O(\alpha_s)$ correction to the QCD factorization formula 
for $B\to \pi\pi$. The method suggested here can be applied
to matrix elements of different topologies and operators and 
to various other $B$-decay channels.   
I also comment on the earlier applications of QCD sum rules 
to nonleptonic decays of heavy mesons.\\

hep-ph/0012271 

\end{abstract}

\vspace*{\fill}

 \noindent $^a${\small \it On leave from 
Yerevan Physics Institute, 375036 Yerevan, Armenia } \\

\newpage

\section{Introduction} 

Nonleptonic two-body $B$ decays such as $B\to \pi \pi$ 
\cite{Bpipi} are very  important 
tools for studying the mechanism of CP-violation (recent overviews of 
theory and experimental perspectives can be found 
in Refs.~\cite{BaBar,Balllhc}). 
In order to fully explore the data on these decays 
one needs reliable theoretical predictions on  
hadronic matrix elements of the operators entering 
the effective weak Hamiltonian
\be
H_W= \frac{G_F}{\sqrt{2}}V_{ub}V^*_{ud}\left\{ 
\left( c_1(\mu)+ \frac{c_2(\mu)}{3}\right)O_1(\mu) + 2c_2(\mu)\widetilde{O}_1(\mu)
+... \right\}\,. 
\label{H}
\ee
The operators relevant for $B\to \pi\pi$ are   
\be
O_1=(\bar{d}\Gamma_\mu u)(\bar{u}\Gamma^\mu b)\,,~~~
\widetilde{O}_1=(\bar{d}\Gamma_\mu \frac{\lambda^a}2u)(\bar{u}\Gamma^\mu
\frac{\lambda^a}2 b)\,,
\label{01}
\ee 
and the penguin operators denoted by ellipses 
(for a comprehensive review of the effective Hamiltonian see, e.g., 
Ref.~\cite{Hrev}). In the above $\Gamma_\mu=\gamma_\mu(1-\gamma_5)$,
$Tr(\lambda^a\lambda^b) = 2\delta^{ab}$,
and $O_2=(\bar{u}\Gamma_\mu u)(\bar{d}\Gamma^\mu b)\,$ 
has been Fierz transformed:
\be
O_2=\frac{1}3 O_1+ 2\widetilde{O}_1\,.  
\ee
In Eq.(\ref{H}), $\mu \sim m_b$ is the normalization scale, 
and all hard gluon effects at distances shorter than $1/\mu$  
are taken into account in perturbative QCD in a form of the 
Wilson coefficients $c_{1,2,...}$. 
The remaining contributions at distances $>1/\mu$  have to be included 
in the hadronic matrix elements, such as $\langle\pi\pi|O_i|
B\rangle$.
Even an approximate calculation of these matrix elements  
is a very difficult problem for the available QCD methods, 
in particular, for the lattice QCD. 
Additional complications arise due to the fact that, as a rule, 
several quark topologies (emission, annihilation, penguin, etc.) 
contribute to each  matrix element 
(a general analysis can be found in Ref.~\cite{BurasSilv}).

The method of QCD light-cone sum rules (LCSR) \cite{lcsr,BF1,CZ} 
offers a possibility to 
calculate exclusive hadronic amplitudes by matching them, 
via dispersion relations and quark-hadron duality, with the correlation 
functions of quark currents. These functions are evaluated in 
the spacelike region using OPE near the light-cone. 
Among  other applications, various heavy-to-light hadronic form 
factors have been obtained  by this method. Importantly, the
predicted form factors include both soft (end-point)  and hard 
rescattering contributions and  reveal a regular $1/m_b$ expansion 
(for reviews of the method and applications, 
see Refs.~\cite{KRHF,CK}). In this paper a new LCSR technique 
is suggested to calculate hadronic matrix elements for 
two-body nonleptonic $B$ decays. The sum rule is obtained from the 
light-cone expansion of a three-point correlation function.  
As a study case we consider the $\bar{B}^0_d\to \pi^+\pi^-$ 
channel.

One of the main objectives of this work is to quantitatively assess 
the factorization approximation
for $B\to \pi\pi$. Recently, the QCD factorization approach 
has been developed \cite{BBNS99} for two-body nonleptonic $B$ decays. 
It was shown that in the limit of 
the heavy $b$ quark mass the decay  
amplitudes can be treated in QCD using the framework of 
perturbative factorization for exclusive processes
\cite{hard}, with the heavy mass playing the same
role as the large momentum transfer. In the limit $m_b \to \infty$ ,
the effects violating factorization in $B\to \pi\pi$ are
suppressed either by powers of $\alpha_s(m_b)$ or by powers
of $\Lambda_{QCD}/m_b$. 

The $B\to \pi\pi$ amplitude obtained below from LCSR has a very similar 
structure to the one predicted from QCD factorization.  
It contains a factorizable
part, where the $B\to \pi$ form factor is itself given by 
LCSR, including both soft and hard 
contributions.  Furthermore, and it is the main result of 
this paper, the sum rule approach allows one to calculate 
the effects of soft gluons that violate factorization.
In LCSR, these effects are generated by higher-twist contributions 
to the correlation function and are therefore under control.    
Hard nonfactorizable effects correspond to $O(\alpha_s)$ 
corrections to the same correlation function, but their estimate 
demands technically complicated two-loop calculations. 
The new method suggested here avoids 
certain problems of the earlier applications of QCD sum rules 
to nonleptonic $D$ and $B$ decays \cite{BS87,BS93,nonfactSR}.
Moreover, we demonstrate that while the use of the light-cone OPE
yields a $1/m_b$ behavior of soft higher-twist effects, 
the expansion in  local operators employed before 
fails to reproduce the $1/m_b$ suppression.

This paper contains a description 
of the method and the first results for 
$B \to \pi\pi$. We confine ourselves to the 
contributions of the current-current operators $O_{1,2}$
in the emission topology. Analysis of 
other topologies (annihilation, penguin) and operators 
(quark and gluon penguins),  as well as applications of the same
method to other $B$ decay channels 
will be published elsewhere.  

The content of this paper is as follows. 
Section 2 contains a derivation of the LCSR for $B\to \pi\pi$. 
In Section 3 the factorizable part of the hadronic matrix 
element $\langle \pi^- \pi^+ |O_1 |\bar{B}^0_d \rangle$
is reproduced and the soft-gluon nonfactorizable effects determined by 
$\langle \pi^- \pi^+ |\widetilde{O}_1 |\bar{B}^0_d \rangle$ are
calculated. 
Using these results, in Section 4 we discuss the 
influence of the soft nonfactorizable correction 
on the QCD factorization formula. Section 5 is devoted to 
the comments on the earlier use of QCD sum rules for 
nonleptonic decays of heavy mesons.
A summary is given in Section 6.

\section{The method}

In what follows, we concentrate on the $\bar{B}^0_d \to \pi^+\pi^-$ 
hadronic matrix elements of the operators $O_{1}$ and $\widetilde{O}_{1}$. 
As a starting object for the derivation  of LCSR we choose the following  
vacuum-pion correlation function:
\be
F^{(O)}_\alpha (p,q,k)=
-\int d^4x\, e^{i(p-q)x} \int d^4y\, e^{i(p-k)y}
\langle\, 0 \mid T\{j^{(\pi)}_{\alpha 5}(y)O(0)j^{(B)}_5(x)\}\mid \pi^-(q)
\rangle\,,
\label{corr}
\ee
where $j^{(\pi)}_{\alpha5}=\bar{u}\gamma_\alpha \gamma_5 d $ 
and $j^{(B)}_5 = m_b \bar{b}i \gamma_5 d$ are the quark currents interpolating 
$\pi$ and $B$ mesons, respectively, and $O$ is either $O_1$ or 
$\widetilde{O}_1$. The correlator (\ref{corr}) is a function of 
three independent momenta, which are chosen to be $q$, $p-k$ and $k$. 
The pion is on-shell, $q^2=m_\pi^2$. 
We will work in the chiral limit
and set $m_\pi=0$ everywhere, unless there is a chirally enhanced 
combination $m_\pi^2/(m_u+m_d)$.

The four-momentum $k\neq 0$ introduced in Eq.~(\ref{corr})
is unphysical because in reality there is no external momentum flow 
from the weak operator vertex. This momentum is added deliberately,
 otherwise in the correlator the quark states before and after 
the $b$-quark decay will have one and the 
same four-momentum, and the dispersion relation in the 
variable of this momentum squared
will contain, below the pole of the ground state $B$ meson, 
a continuum  of light ``parasitic'' contributions. At $k\neq 0$, 
the momentum $p-q$ in the $B$ channel is independent of  
$P\equiv p-k-q $, the total momentum of the state 
formed after the $b$ quark decay.  
As we shall see below, the procedure is designed in such a way that in 
the final dispersion relation the 4-momentum $k$ automatically vanishes 
in the  $B\to \pi\pi$ ground-state contribution.

The decomposition of the correlation function (\ref{corr})
in independent momenta is straightforward and contains four
invariant amplitudes:
\be
F_\alpha^{(O)}= 
(p-k)_\alpha F^{(O)} 
+ q_\alpha \widetilde{F}_1 ^{(O)} 
+ k_\alpha\widetilde{F}_2^{(O)} 
+ \epsilon_{\alpha\beta\lambda\rho}q^\beta p^\lambda k^\rho
\widetilde{F}_3^{(O)}\,. 
\label{decompos}
\ee
In what follows only the amplitude 
$F^{(O)}$ is relevant. 
The correlation function (\ref{corr}) at $k\neq $ 0 is, in fact, 
a $2\to 2$  scattering amplitude. Therefore, counting  
independent kinematical invariants, one has to add 
two additional variables $P^2$ and $p^2$ to the  
four external momenta squared $q^2=0$, $(p-q)^2$, $k^2$, and $(p-k)^2$.

The correlation function is calculated in QCD by expanding the T-product
of three operators, two currents and $O$, near the light-cone 
$x^2\sim y^2\sim (x-y)^2\sim 0$. For this expansion 
to be valid, the kinematical region should be 
carefully chosen, in order to stay far away from hadronic 
thresholds in both channels of $\pi$ and $B$ currents.  
Consequently, the external momenta squared $(p-q)^2$ 
and $(p-k)^2$, as well as the kinematical invariant $P^2$ have to 
be taken spacelike and large. Simultaneously,
to simplify calculations, it is possible  
to put the remaining external momentum squared $k^2$ and 
the kinematical invariant $p^2$ to zero having in mind that  
hadronic states in the channels with four-momenta $k$
and $p$ contain heavy $b$ quark and, therefore, the 
corresponding thresholds are located far enough from the 
zero points.  To summarize, our choice
of the kinematical region in Eq.~(\ref{corr}) is 
\be
q^2= p^2=k^2=0\,,~~~
|(p-k)^2|\sim |(p-q)^2|\sim |P^2| \gg \Lambda^2_{QCD}\,.
\label{region}
\ee

In this region the light-cone expansion of 
the correlation function (\ref{corr}) is applicable 
and the result is obtained in terms of hard scattering amplitudes convoluted
with the pion light-cone distribution amplitudes of different twist.
The actual calculation is described in the next section. At this point,
it suffices to represent the result of 
the QCD calculation in a generic form of a 
dispersion relation in the variable $(p-k)^2$:
\be
F^{(O)}_{QCD}((p-k)^2,(p-q)^2,P^2)=
\frac1\pi\int\limits_{0}^{\infty}ds~
\frac{\mbox{Im} F^{(O)}_{QCD}(s,(p-q)^2,P^2)}{s-(p-k)^2}\,.
\label{dispQCD}
\ee

To proceed with the derivation, we obtain a corresponding hadronic 
dispersion relation, inserting in Eq.~(\ref{corr}) a complete 
set of intermediate states with the pion quantum numbers:
\be
F^{(O)}((p-k)^2,(p-q)^2,P^2)= 
\frac{ if_\pi 
\Pi^{(O)}_{\pi\pi}((p-q)^2,P^2)}{-(p-k)^2}
+\int\limits_{s_h^{(\pi)}}^{\infty} ds~\frac{\rho_h^{(\pi)}(s,(p-q)^2,P^2)}{s-(p-k)^2}\,,
\label{disp}
\ee
where the lowest, one-pion contribution contains 
the pion decay constant 
\be
\langle 0 |\bar{u}\gamma_\alpha \gamma_5 d |\pi(p-k)\rangle 
=if_\pi (p-k)_\alpha\,,
\label{fpi}
\ee
and the hadronic matrix element 
\be
\Pi^{(O)}_{\pi\pi}((p-q)^2,P^2)= i\int d^4x\; 
e^{i(p-q)x} \langle \pi^-(p-k)|T\{ O(0) j^{(B)}_5(x)\} | \pi^-(q) \rangle\,,
\label{twopion}
\ee
which itself is a two-point correlation function. The spectral function 
$\rho_h^{(\pi)}$ in Eq.~(\ref{disp}) accounts for the 
contributions of excited states and continuum in the pion channel,
$s_h^{(\pi)}$ being the lowest threshold.
Subtraction terms in both dispersion relations 
(\ref{dispQCD}) and (\ref{disp}) are omitted in anticipation of the Borel 
transformation which will  remove them anyway.
Physically, the two-point function   $ \Pi^{(O)}_{\pi\pi}$  defined
in Eq.~(\ref{twopion}) describes a   
scattering process. In more detail, the pion undergoes a
deep elastic scattering due to a combined action 
of the current  $j^{(B)}_5$ and operator $O$, 
with a total spacelike 
momentum transfer $P^2$. 
To obtain $\Pi_{\pi\pi}^{(O)}$ we follow the usual derivation procedure
of QCD sum rules \cite{SVZ} and match the dispersion relation 
(\ref{disp}) with the 
result (\ref{dispQCD}) of the light-cone expansion 
at large $|(p-k)^2|$. Furthermore, the integral over $\rho_h^{(\pi)}$
is approximated using the quark-hadron duality:
\be 
\int\limits_{s_h^{(\pi)}}^{\infty} ds~\frac{\rho_h^{(\pi)}(s,
(p-q)^2,P^2)}{s-(p-k)^2} = 
\frac1\pi\int\limits_{s_0^\pi}^{\infty}ds~
\frac{\mbox{Im} F^{(O)}_{QCD}(s,(p-q)^2,P^2)}{s-(p-k)^2}\,,
\ee
where $s_0^{\pi}$ is an effective threshold parameter. 
After the Borel transformation in the variable $(p-k)^2$ the result reads 
\be
\Pi^{(O)}_{\pi\pi}((p-q)^2,P^2)= 
-\frac{i}{\pi f_\pi}\int\limits_{0}^{s_0^{\pi}}ds~
e^{-s/M^2}\mbox{Im}_{s} F^{(O)}_{QCD}(s,(p-q)^2,P^2)\,.
\label{respipi}
\ee

Having at hand an expression for 
$\Pi_{\pi\pi}^{(O)}((p-q)^2,P^2)$ 
valid at large spacelike $P^2$,  we perform an analytic
continuation to large timelike values of $P^2$, keeping the 
variable $(p-q)^2$ fixed. This procedure is very 
similar to the  evaluation  of the timelike asymptotics 
of the pion electromagnetic form factor from a 
QCD calculation at large spacelike 
momentum transfer. For $\Pi_{\pi\pi}^{(O)}$  a  natural and 
convenient point of the analytic continuation
is $P^2=m_B^2$ where the invariant mass of two pions is equal to 
the $B$ meson mass which is sufficiently large, 
$m_B\sim m_b \gg \Lambda_{QCD}$. The analytic continuation 
of Eqs.~(\ref{twopion}) and
(\ref{respipi}) yields a relation 
\bea
\Pi^{(O)}_{\pi\pi}((p-q)^2,m_B^2)= i\int d^4x~ 
e^{i(p-q)x} \langle \pi^-(p-k)\pi^+(-q) | T\{O(0) j^{(B)}_5(x)\} | 0\rangle
\nonumber
\\
=\frac{-i}{\pi f_\pi}\int\limits_{0}^{s_0^{\pi}}ds~
e^{-s/M^2}\mbox{Im}_s F^{(O)}_{QCD}(s,(p-q)^2,m_B^2)
\label{timelike}
\eea
for the hadronic matrix element  
which is crossing-related to the one defined in Eq.~(\ref{twopion})
and corresponds to an  amplitude of two-pion production 
with a large invariant mass $m_B^2$. This process is very 
similar to $\gamma^* \to \pi^+ \pi^-$ but
has a more complicated underlying quark-current structure.
Note also that the analytic continuation
to timelike $P^2$ can generate in 
$\Pi^{(O)}_{\pi\pi}((p-q)^2,m_B^2)$ a complex phase, which
should be identified, within the accuracy of our calculation, with 
the phase of the final-state rescattering of two pions.  

The next step in the derivation procedure employs 
analytic properties of the two-pion amplitude 
(\ref{timelike}) in the remaining spacelike variable $(p-q)^2$. 
Inserting in Eq.~(\ref{timelike}) a complete
set of hadronic states with the $B$ meson quantum numbers 
one obtains the following dispersion relation:  
\be
\Pi^{(O)}_{\pi\pi}((p-q)^2,m_B^2)= 
\frac{f_Bm_B^2
\langle \pi^-(p)\pi^+(-q) |O |B(p-q) \rangle }{m_B^2-(p-q)^2} 
+\int\limits_{s_h^{B}}^{\infty} ds'\frac{\rho_h^{(B)}(s')}{s'-(p-q)^2}\,,
\label{disp2}
\ee
where the standard definition of the $B$-meson decay constant 
$
\langle B|j_5 |0 \rangle =m_B^2 f_B 
$
is used.
Note that in the ground-state contribution 
the auxiliary momentum $k$ vanishes, due to simultaneous 
conditions $(p-q-k)^2=m_B^2$ and $(p-q)^2=m_B^2$, so that  
the on-shell $B\to \pi \pi$ matrix element of the operator $O$ 
is recovered. In the spectral density of excited and continuum states 
$k$ remains nonzero and depends on the value of  $s'>m_B^2$.
At large $|(p-q)^2|$ the relation (\ref{disp2}) is matched with 
the result of the QCD calculation given by the r.h.s. of Eq.~(\ref{timelike}) 
which is rewritten in a form of a 
dispersion relation:
\bea
\Pi^{(O)}_{\pi\pi}((p-q)^2,m_B^2)
=\frac{-i}{\pi^2f_\pi}\int\limits_{0}^{s_0^{\pi}}\!ds
\,e^{-s/M^2}\!\!\int\limits_{m_b^2}^{R(s,m_b^2,m_B^2)} 
\frac{ds'}{s'-(p-q)^2}\;\mbox{Im}_{s'} 
\mbox{Im}_s F^{(O)}_{QCD}(s,s',m_B^2).
\label{pipi2}
\eea
The upper limit $R$ of the integration in $s'$ depends, in general,  
on $s$ and on $P^2=m_B^2$. At this point
one again makes use of quark-hadron duality and approximates 
the dispersion integral over the spectral density $\rho^{(B)}_h$ 
by the $s'\geq s_0^B$ part of the dispersion integral 
(\ref{pipi2}), where $s_0^B$ is the effective threshold in the $B$ channel. 
Finally, the second Borel transformation with respect 
to the variable $(p-q)^2$  is performed.
The resulting LCSR for the $\bar{B}^0_d \to \pi^+ \pi^-$ matrix element 
of the operator $O$ reads:
\bea
&&A^{(O)}( \bar{B}^0_d \to \pi^+ \pi^-) 
\equiv \langle \pi^-(p)\pi^+(-q) |O |B(p-q) \rangle
\nonumber
\\
&=&\frac{-i}{\pi^2f_\pi f_B m_B^2} \int\limits_{0}^{s_0^{\pi}}\!ds~
e^{-s/M^2}\!\!\int\limits_{m_b^2}^{\bar{R}(s,m_b^2,m_B^2,s_0^B)} 
\!ds'\;e^{(m_B^2-s')/M'^2}\mbox{Im}_{s'}\, 
\mbox{Im}_s F^{(O)}_{QCD}(s,s',m_B^2)\,,
\eea
where $\bar{R}$ is the upper limit after the duality subtraction in 
the dispersion integral.

\section{$\bar{B}^0_d \to \pi^+ \pi^-$ hadronic 
matrix elements}

\subsection{Factorizable part}

The remaining task is to calculate $F^{(O)}_{QCD}$ for 
$O=O_1,\widetilde{O}_1$. 
At the diagrammatical level, there are four topologically 
different contributions to  the correlation function (\ref{corr}),
corresponding to four possible
combinations of $\bar{u}$ and $d$ quark-field operators in the 
pion  distribution amplitude 
$\langle 0 \mid \bar{u}_\alpha(z_1)d_\beta(z_2)\mid \pi^-\rangle$,
where $z_{1}=0$ or $y$, 
and $z_2=x$ or $y$ ($ \alpha, \beta$ are the spinor indices). 
Drawing the quark diagrams we find that 
each contribution yields a $B\to \pi\pi$ 
matrix element with a certain quark topology: 
emission ($z_1=0$, $z_2=x$), 
annihilation ($z_1=0$, $z_2=y$), penguin ($z_1=y$, $z_2=x$)
and penguin annihilation ($z_1=z_2=y$).
In what follows we will concentrate 
on the emission topology adding a subscript $E$
to the corresponding quantities. Some of the  
diagrams contributing to $F_{\alpha,E}^{(O)}$ are shown in Figs.~1,2.  
In the case of the operator $O_1$ there are diagrams, where the quarks 
belonging to the heavy-light currents do not 
interact with the quarks of the light-quark currents. 
The leading-order diagram of this type is shown in Fig.~1a.
To single out the contribution of this and all other ``factorizable'' diagrams, 
we insert an intermediate vacuum state between the weak currents 
of the operator $O_1$. Eq.~(\ref{corr}) is then converted 
into a product of two disconnected 
two-point correlation functions: 
\bea
F^{(O_1)}_{\alpha E} (p,q,k)=
\Bigg(i\!\int d^4y\; e^{i(p-k)y}
\langle\, 0 \mid T\{j^{(\pi)}_{\alpha5}(y)\bar{d}(0)\gamma_\mu\gamma_5 u(0) \}\mid 0\rangle
\Bigg)
\nonumber
\\
\times \Bigg(i\!\int d^4x\; e^{i(p-q)x} 
\langle\, 0 \mid T\{\bar{u}(0)\gamma^\mu b(0)j^{(B)}_5(x)\}\mid \pi^-(q)\rangle
\Bigg )\,,
\label{corrE}
\eea
where, due to spin-parity conservation, 
$\Gamma_\mu\to \gamma_\mu\gamma_5$ ($\Gamma_\mu\to \gamma_\mu$) 
in the first (second) weak current.  
To proceed further, we insert complete sets of intermediate states  
with the pion and $B$-meson quantum numbers in the first and 
second correlators in Eq.~(\ref{corrE}), respectively. After that,
the matrix element $\langle \pi^-(p)\pi^+(-q) |O |B(p-q) \rangle$  
is extracted in the usual factorized form: 
\be 
 A^{(O_1)}_{E}( \bar{B}^0_d\to\pi^+\pi^-) = i f_\pi f_{B\pi}^+(0)m_B^2\,,
\label{Afact}
\ee
where $f^+_{B\pi}$ is the $B\to \pi$ form factor defined as
\be
\langle \pi(q)| \bar u \gamma_\mu b| B(p+q)\rangle =
2f^+_{B\pi}(p^2) q_\mu + (f^+_{B\pi}(p^2)+ f^-_{B\pi}(p^2))p_\mu\,.
\label{Bpi}
\ee
and taken at $p^2=m_\pi^2\simeq 0$. Correction of $O(m_\pi^2/m_B^2)$ 
to Eq.~(\ref{Afact}) is neglected in the chiral limit adopted here.

Within  the sum rule method, it is possible 
to calculate the r.h.s. of Eq.~(\ref{Afact}) 
from Eq.~(\ref{corrE}), with a certain accuracy. 
In leading-order, $F^{(O_1)}_{\alpha E}$ is given by the 
diagram in Fig.~1a. The answer is easily obtained   by forming the pion 
distribution amplitude$~~~$ 
$\langle 0 \!\mid \bar{u}_\alpha(0)d_\beta(x) \mid \pi^-\rangle$,
contracting the remaining quark fields in both 
correlation functions and using 
free-quark propagators. Integrating over the coordinates $x,y$ 
we obtain:
\be
F^{(O_1)}_{E}= -\frac{m_b^2f_\pi}{4\pi^2}q\cdot(p-k)
\ln\left(-(p-k)^2\right)
\int\limits_0^1
du \frac{\varphi_\pi(u)}{m_b^2-(p-q(1-u))^2}\,.
\label{Ftw2}
\ee
In the above, the logarithm corresponds to the bare-loop approximation 
of the first (vacuum-vacuum) correlation function
in Eq.~(\ref{corrE}) \footnote{The divergent part of the 
loop  vanishes after the Borel transformation 
and is therefore omitted from Eq.~(\ref{Ftw2}).}, 
whereas the integral represents the second (vacuum-pion) 
correlation function. In the latter function, only the leading twist-2 part 
is retained,  and the pion distribution
amplitude $\varphi_\pi$ is defined in a standard way:
\be  
\langle 0|\bar{u}(0)\gamma_\mu\gamma_5 d(x)|\pi(q)\rangle=
iq_\mu f_\pi\int_0^1du\,e^{-iu q\cdot x}
\varphi_\pi (u,\mu)\,,
\label{piontw2}
\ee
where
\be
\varphi_\pi(u,\mu) = 6u(1-u) \left[1+ \sum_{n=2,4,..} 
a_{n}(\mu)C_{n}^{3/2}(2u-1)\right]\,, 
\label{gegenb}
\ee
$C_{n}^{3/2}$ are the Gegenbauer polynomials and 
$a_{n}$ are multiplicatively renormalizable coefficients 
depending on the normalization scale $\mu$.
Following  the procedure described in Sect. 2 and applying,
subsequently, duality with Borel transformation in the pion channel,    
analytic continuation $P^2 \to m_B^2$ and duality with Borel
transformation in the $B$-meson channel, one obtains:
\bea
 A^{(O_1)}_{E,tw2}( \bar{B}^0_d\to\pi^+\pi^-)&=& im_B^2
\Bigg(\frac{1}{4\pi^2f_\pi}\int\limits_0^{s_0^\pi}
ds\left(1-\frac{s}{m_B^2}\right)e^{-s/M^2}
\Bigg)
\nonumber
\\
&&\times 
\Bigg(\frac{m_b^2f_\pi}{2f_Bm_B^2}\int\limits_{u_0^B}^1 \frac{du}u~
e^{\,m_B^2/M'^2-m_b^2/uM'^2 }\varphi_\pi(u,\mu_b)\Bigg)\,,
\label{Afacttw2}
\eea
where $u_0^B= m_b^2/s_0^B$ and $\mu_b$ is an appropriate scale.
In this expression, the first bracket, up to a small $O(s/m_B^2) 
\sim O(s_0^\pi/m_B^2)$ correction,  is approximately 
equal to $f_\pi$, because the integral over $s$ gives the quark-loop 
(leading order) term of the SVZ sum rule \cite{SVZ} for $f_\pi^2$. 
The second bracket 
in Eq.~(\ref{Afacttw2}) coincides with the leading twist 2 term in the
LCSR \cite{CZ} for the form factor $f^+_{B\pi}$. 
Thus, the factorizable matrix element (\ref{Afact})
is restored at the leading-order level.  Factorizable 
corrections in both correlation 
functions entering Eq.~(\ref{corrE}) can be added, improving the accuracy 
of Eq.~(\ref{Afacttw2}). 
The LCSR for the form factor $f^+_{B\pi}$ \cite{BKR} contains, 
in addition to the twist 2 term, the  twist 3, 4  contributions 
corresponding to two- and three-particle distribution amplitudes
of the pion. 
The numerically important twist 3 contribution is proportional
to the chirally enhanced coefficient $m_\pi^2/(m_u+m_d)$.
In addition, the $O(\alpha_s)$ radiative correction 
to the twist 2 part has been calculated \cite{KRWY,Bagan}.
A certain part of this correction \cite{Bagan} corresponds to the 
hard rescattering mechanism. Summarizing, the 
hadronic matrix element of the operator $O_1$  in the emission topology 
is given by Eq.~(\ref{Afact}) where $f^+_{B\pi}$ itself is calculated 
from LCSR with an available accuracy. 

%%%%%%%%% Figure:Bpipidiag%%%%%%%%%%%%%%%%%
\begin{figure}
%\rule{5cm}{0.2mm}\hfill\rule{5cm}{0.2mm}
%\vskip 0.5cm
%\rule{5cm}{0.2mm}\hfill\rule{5cm}{0.2mm}
\hspace{1cm}
\psfig{figure=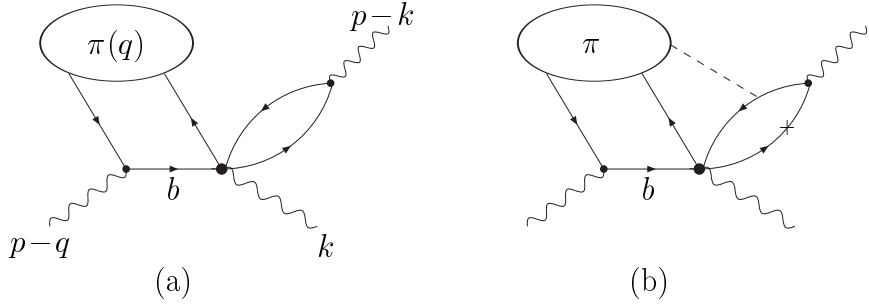,height=2.2in}
\caption{ {\it Diagrams corresponding (a) to the leading-order
of the correlation function (\ref{corr}) for $O=O_1$;
(b) to the higher-twist soft-gluon nonfactorizable contribution
for $O=\widetilde{O}_1$. 
Solid, dashed and wavy lines represent quarks, gluons, and 
external momenta, respectively. Thick points denote the 
weak interaction vertices, and ovals the pion distribution amplitudes. 
The cross indicates the point of 
gluon emission in the second similar diagram.}}
\end{figure}
%%%%%%%%%%%%%%%%%%%%%%%%%%%%

\subsection{Nonfactorizable effects}

In the case of the operator $O_1$, nonfactorizable corrections to the 
correlation function (\ref{corr}) emerge 
at a two-gluon level and are neglected here. 
Calculation of the matrix element $\langle \pi^-\pi^+|\widetilde{O}_1 |B \rangle$
brings nonfactorizable effects at a one-gluon 
level into the game. The relevant correlation
function $F_\alpha^{(\widetilde{O}_1)}$ receives contributions of
hard gluon exchanges which, at  $O(\alpha_s)$, correspond to the diagrams 
in Fig.2. These two-loop diagrams can be calculated in QCD perturbation
theory, but this task is beyond the scope of this paper. 
In future, it is important  to obtain 
this part of the correlation function, not only for the sake of 
improving the accuracy of the LCSR, but also for 
other important reasons. First, 
at $O(\alpha_s)$, the matrix element should partially 
compensate the scale-dependence 
of the short-distance coefficients $c_{1,2}$ in the effective 
weak Hamiltonian. 
Second, the analytic continuation of these two-loop contributions in $P^2$ 
will generate an imaginary part, yielding  
a final-state rescattering phase in the decay amplitude. 
Third, it will be interesting to compare the  $O(\alpha_s)$ part of the LCSR 
with the hard nonfactorizable corrections obtained 
in the QCD factorization approach \cite{BBNS99}. 
In fact, there is a certain correspondence 
between the diagrams in Fig. 2 and the quark diagrams contributing to 
the hard-scattering kernels of the QCD factorization.
More specifically, the diagrams in Fig. 2a,b, where the hard  
gluon is exchanged between the quarks participating in the weak interaction, 
are of the non-spectator type, whereas the diagrams in Fig. 2c correspond to 
the hard spectator contribution.  In particular, it would be 
instructive to check if 
at $m_b\to \infty$ the form factor $f^+_{B\pi}$ factorizes out in the
non-spectator part of LCSR, as it does in the QCD factorization formula. 
\footnote{Differences may arise
due to the fact that in the LCSR approach one of the 
pions and $B$ meson are represented by the integrals over quark spectral
densities. In these integrals,  quark transverse momenta up 
to $O(s_0^\pi)$ and $O(s_0^B)$, respectively, are taken into account, 
whereas in QCD factorization 
these hadrons are described by collinear 
light-cone distribution amplitudes.}
%%%%%%%%% Figure:Bpipihard%%%%%%%%%%%%%%%%%
\begin{figure}[t]
%\rule{5cm}{0.2mm}\hfill\rule{5cm}{0.2mm}
%\vskip 0.5cm
%\rule{5cm}{0.2mm}\hfill\rule{5cm}{0.2mm}
\hspace{1cm}
\psfig{figure=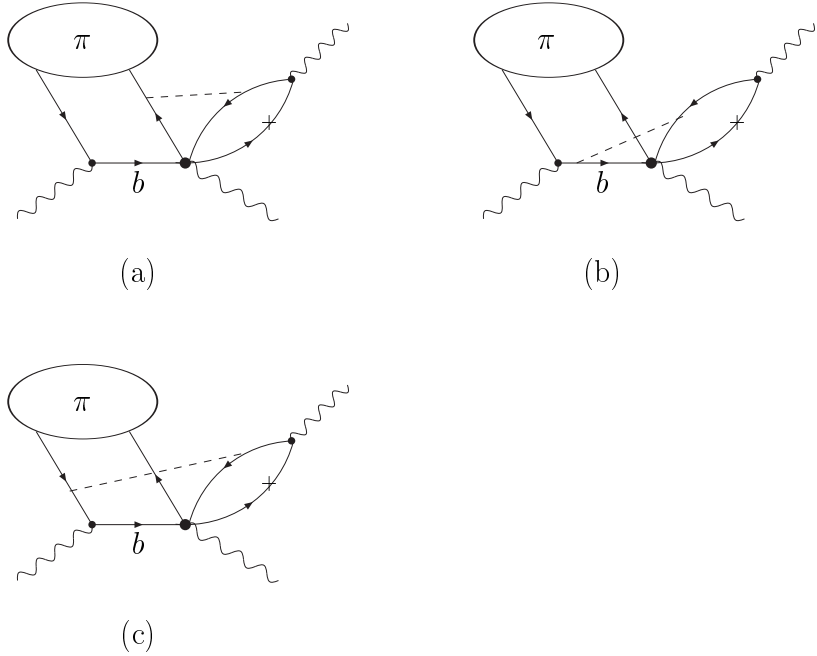,height=4.0in}
\caption{{\it 
Diagrams corresponding to the 
$O(\alpha_s)$ nonfactorizable contributions 
to the correlation function (\ref{corr})
for $O=\widetilde{O}_1$.}}
\end{figure}
%%%%%%%%%%%%%%%%%%%%%%%%%%%%

There is another class
of nonfactorizable effects contributing to $F^{(\widetilde{O}_1)}$,
with on-shell gluons  emitted from the quarks of the 
pion current  and absorbed  in the pion 
distribution amplitude, as shown in Fig. 1b. 
In terms of the light-cone expansion these contributions are of a higher 
twist, starting from twist 3.  
The diagrams in Fig. 1b are calculated employing the light-cone expansion 
of the quark propagator \cite{BB89}: 
\bea 
S(x,0)&\equiv& -i\langle 0 | T \{q(x) \bar{q}(0)\}| 0 \rangle 
= 
\frac{\Gamma(d/2)\not\!x}{2\pi^2(-x^2)^{d/2}}  
\nonumber
\\
&+&
\frac{\Gamma(d/2-1)}{16\pi^2(-x^2)^{d/2-1}}\int\limits_0^1dv 
\Big\{(1-v)\not\!x \sigma_{\mu \nu} G^{\mu\nu}(vx) + 
v\sigma_{\mu \nu} G^{\mu\nu}(vx)\!\not\! x \Big\} + ...,
\label{prop}
\eea
where $ G_{\mu\nu}= g_sG_{\mu\nu}^a(\lambda^a/2)$, 
and $d$ is the space-time dimension.
In the above, the fixed-point gauge for the gluon field is adopted and 
only the terms proportional to the one gluon-field strength 
are shown, which is  the accuracy we need here.

The relevant quark-antiquark-gluon distribution amplitudes 
are defined by the matrix elements \cite{BF} 
\bea
%\lefteqn{ 
&&\langle 0 |\bar{u}(0)
\sigma_{\mu\nu}\gamma_5 G_{\alpha\beta}(vy) d(x)| \pi(q) \rangle 
%=}
=if_{3\pi}[(q_\alpha q_\mu g_{\beta\nu}-q_\beta q_\mu g_{\alpha\nu})
\nonumber\\
&&-(q_\alpha q_\nu g_{\beta\mu}-q_\beta q_\nu g_{\alpha\mu})]
\int{\cal D}\alpha_i\,\varphi_{3\pi}(\alpha_i,\mu)e^{-iq(x\alpha_1+yv\alpha_3)}\,,
\label{tw3matr}
\eea
\bea 
&&\langle 0| \bar u(0)i\gamma_\mu \widetilde G_{\alpha\beta}(vy)
d(x) |\pi(q)\rangle = q_\mu \frac{q_\alpha x_\beta-q_\beta x_\alpha}{qx}
f_\pi \int\! {\cal D}\alpha_i 
\widetilde \varphi_\parallel(\alpha_i,\mu) e^{-iq(x\alpha_1+yv\alpha_3)}
\nonumber \\
&&+ ( g^\perp_{\mu\alpha}q_\beta  - g^\perp_{\mu\beta}q_\alpha)
f_\pi \int\! {\cal D}\alpha_i 
\widetilde\varphi_\perp(\alpha_i,\mu) e^{-iq(x\alpha_1+yv\alpha_3)},
\label{tw4matr}
\eea 
and the one analogous to Eq.~(\ref{tw4matr})
with $i\gamma_\mu \to \gamma_\mu\gamma_5 $,
$\widetilde{G}_{\alpha\beta} \to  G_{\alpha\beta}$ and 
$\widetilde{\varphi}_{\parallel,\perp} \to \varphi_{\parallel,\perp}$.
In the above, 
$\widetilde G_{\alpha\beta}= \frac 12 
\epsilon_{\alpha\beta\rho\lambda}G^{\rho\lambda}$, 
${\cal D} \alpha_i = d\alpha_1d\alpha_2d\alpha_3
\delta \left(1- \alpha_1 - \alpha_2- \alpha_3 \right)$, and
$
g_{\alpha\beta}^\perp = g_{\alpha\beta} - 
(x_\alpha q_\beta + x_\beta q_\alpha)/qx \,.
$
The pion distribution amplitude 
$\varphi_{3\pi}$  is  of twist 3, whereas 
$\widetilde{\varphi}_{\parallel,\perp}$ and  $\varphi_{\parallel,\perp}$ 
are of twist 4. In what follows we suppress the $\mu$ dependence for
brevity.

A straightforward calculation of the two diagrams in Fig.~1b yields
the following answer for the twist 3 contribution:
\bea
F^{(\widetilde{O}_1)}_{E,tw3}=\frac{m_b f_{3\pi}}{4\pi^2}
\int\limits_0^1 dv\int {\cal D}\alpha_i ~
\frac{\varphi_{3\pi}(\alpha_i)}{(m_b^2-(p-q)^2(1-\alpha_1))(-P^2v\alpha_3-
(p-k)^2(1-v\alpha_3))}
\nonumber
\\
\left[(2-v)(q\cdot k) +2(1-v)q\cdot(p-k)\right] (q\cdot(p-k))\,,
\label{tw3F}
\eea
which is then easily transformed into a form of the dispersion integral 
(\ref{dispQCD}):
\bea
 F^{(\widetilde{O}_1)}_{E,~tw3}= -\frac{m_b f_{3\pi}}{16\pi^2}
\int\limits_0^\infty \frac{ds}{s-(p-k)^2}
\int\limits_{s/(s-P^2)}^{1}\frac{du}{m_b^2-(p-q)^2u}
\nonumber
\\
\times\int\limits_{s/(s-P^2)}^{u}\frac{dv}{v^2}
\,\varphi_{3\pi}(1-u, u-v, v)
\left[s+\frac{m_b^2}{u}\left(2v-\frac{s}{s-P^2} \right)\right]\,,
\label{corrtw3}
\eea
For simplicity, all terms vanishing after Borel transformations 
are omitted in this expression.
In the part of Eq.~(\ref{corrtw3}) which is dual to  a pion, 
the upper limit of the first integral is $s_0^\pi \ll |P^2|\sim m_b^2$.
Hence, the resulting expression for the two-pion matrix 
element $\Pi^{(\widetilde{O}_1)}_{\pi\pi}((p-q)^2,P^2)$ 
can be simplified by expanding it in powers of $s/P^2$, 
and neglecting  very small $O(s/P^2) \sim O(s_0^\pi/P^2)$ corrections:
\bea
 \Pi^{(\widetilde{O}_1)}_{\pi\pi,~tw3}((p-q)^2,P^2)
=\frac{im_b^3 f_{3\pi}}{8\pi^2f_\pi}
\int\limits_0^{s_0^\pi} ds~e^{-s/M^2}
\int\limits_{0}^{1}\frac{du}{u(m_b^2-(p-q)^2u)}
\nonumber
\\
\times\int\limits_{0}^{u}\frac{dv}{v}
\varphi_{3\pi}(1-u, u-v, v)
\left\{1+ O\left(s_0^\pi/P^2 \right )\right\}\,.
\label{pipitw3}
\eea
The analytic continuation $P^2\to m_B^2$ is then trivial.
The twist 4 contribution to $\Pi^{(\widetilde{O}_1)}_{\pi\pi}$
has a structure similar to Eq.~(\ref{pipitw3}) but a bulky expression 
not shown here for brevity.  
Finally, the LCSR for the $\bar{B}^0_d\to \pi^+\pi^-$ matrix element 
of the operator $\widetilde{O}_1$ is obtained 
applying to $\Pi^{(\widetilde{O}_1)}_{\pi\pi}((p-q)^2,m_B^2)$ the
duality approximation and Borel transformation in the $B$ channel. 
The result is:  
\bea
&&A^{(\widetilde{O}_1)}_{E}( \bar{B}_d^0\to\pi^+\pi^-)= im_B^2
\Bigg(\frac{1}{4\pi^2f_\pi}\int\limits _0^{s_0^\pi}ds~
e^{-s/M^2}\Bigg)\Bigg(\frac{m_b^2}{2f_Bm_B^4}\int\limits_{u_0^B}^1 
\frac{du}{u}e^{\,m_B^2/M'^2-m_b^2/uM'^2}
\nonumber
\\
&&\times\Bigg[
\frac{m_bf_{3\pi}}{u} \! \int\limits_0^u\frac{dv}{v}~
\varphi_{3\pi}(1-u,u-v,v)
+f_\pi\int\limits_0^u \frac{dv}{v}
~\Big [ 3\widetilde{\varphi}_{\perp}(1-u,u-v,v)
\nonumber
\\
&&- \!\left(\frac{m_b^2}{uM^2}-1\right)
\frac{\Phi_1(1-u,v)}{u} \Big ] + \left(\frac{m_b^2}{uM'^2}-\frac{s}{M'^2}-1 \right)\!\!\frac{\Phi_2(u)}{u^2}\Bigg]\Bigg)\!\Bigg\{ 
1+ O(s_0^\pi/m_B^2)\Bigg\},
\label{Atw34}
\eea
where the following definitions are introduced:
\be
\frac{\partial\Phi_1(w,v)}{\partial w}= 
\widetilde{\varphi}_{\perp}(w,1-w-v,v) +
\widetilde{\varphi}_{\parallel}(w,1-w-v,v),~~~ 
\frac{\partial\Phi_2(v)}{\partial v}= \Phi_1(1-v,v). 
\ee
The asymptotic form of the pion distribution amplitudes 
in Eq.~(\ref{Atw34}) is given by \cite{BF} 
\be
\varphi_{3\pi}(\alpha_i)=360 \alpha_1\alpha_2\alpha_3^2,~~
\tilde{\varphi}_\perp (\alpha_i)=10\delta^2\alpha_3^2
(1-\alpha_3),~~ 
\tilde{\varphi}_\parallel (\alpha_i)=
-40\delta^2\alpha_1\alpha_2\alpha_3\,.
\label{DA}
\ee
More complete expressions containing scale-dependent nonasymptotic 
terms can be found in Ref.~\cite{KRHF}.  

A few comments are in order concerning LCSR (\ref{Atw34}). 
Firstly, the magnitude of the nonfactorizable effect 
predicted by this sum rule is determined by the two nonperturbative 
parameters $f_{3\pi}$ and $\delta^2$.   
They are defined by the vacuum-pion matrix elements 
$\langle 0 |\bar{u}
\sigma_{\mu\nu}\gamma_5 g_s G_{\alpha\beta}d|\pi\rangle
$
and 
$ \langle 0 |\bar{u} g_s\widetilde{G}_{\alpha\mu}\gamma^\alpha d|\pi \rangle$,
respectively.
Secondly, at leading order in $s_0^{\pi}/m_B^2$, 
the matrix element (\ref{Atw34}) factorizes 
into $f_\pi$ and an expression which has a typical structure of the 
LCSR for a heavy-light form factor.  
Finally, the analytic continuation in $P^2$ 
does not produce an imaginary part in the nonfactorizable 
matrix element $A^{(\widetilde{O}_1)}_{E}$. This fact can be 
interpreted as an absence of a final-state rescattering phase 
induced by soft gluon exchanges.

\subsection{Heavy quark limit}

The heavy-quark mass behavior of the matrix elements 
(\ref{Afacttw2}) and (\ref{Atw34})
can be easily derived. To this end, 
one  has to substitute in the LCSR  the standard expansions 
of quantities dependent on the heavy quark mass:  
\be
m_B = m_b+\bar{\Lambda}~,~~~ s_0^B = m_b^2 + 2m_b\omega_0 ~,
~~~M'^2= 2m_b\tau~,~~~f_B = \hat{f}_B/\sqrt{m_b}, 
\label{hqet}
\ee
where 
$\bar{\Lambda}$, $\omega_0$, $\tau$, and $\hat{f}_B$ are
$m_b$-independent parameters. One also has  to take into account 
the end-point behavior of the light-cone distribution amplitudes, since
at $m_b\to \infty$, $u_B^0\to 1-\omega_0/m_b$ and only the end-point
regions of the momentum fraction $u$ contribute to Eqs.~(\ref{Afacttw2})
and (\ref{Atw34}). As a result we get:
\bea
A^{(O_1)}_{E,tw2}( \bar{B}^0_d\to\pi^+\pi^-)\sim \sqrt{m_b},~~
A^{(\widetilde{O}_1)}_{E}( \bar{B}^0_d\to\pi^+\pi^-) \sim 
\frac{1}{\sqrt{m_b}}\,.
\eea
Thus, according to LCSR, the higher-twist nonfactorizable effects 
are suppressed by one power of $1/m_b$ with respect to the 
twist 2 factorizable amplitude.

\subsection{ Numerical estimates}

The sum rule (\ref{Atw34}) is derived for a finite $b$-quark mass, 
and it is therefore possible to obtain a numerical estimate of
the nonfactorizable matrix element 
$A^{(\widetilde{O}_1)}_{E}( \bar{B}^0_d\to\pi^+\pi^-)$ 
and compare it with the factorizable approximation (\ref{Afact}).

To specify the input, we take 
$f_\pi=132$ MeV, $s_0^\pi=0.7$ GeV$^2$ and $M^2= 0.5\div 1.2 $ GeV$^2$
for the parameters of the pion channel. 
The value of the effective threshold and the interval of the 
Borel parameter are provided by the two-point QCD sum rule 
\cite{SVZ} for $f_\pi$. For the parameters of the $B$-meson channel
and for the pion distribution amplitudes we adopt 
the same input as the one used in LCSR for the form factor $f_{B\pi}^+$
(for a recent update see Ref.~\cite{KRWWY00}):
$f_B=180\pm 30$ MeV, $m_b=4.7 \mp 0.1$ GeV, $s_0^B= 35\pm 2$ GeV$^2$
(the values obtained from the QCD sum rule for $f_B$), 
$\mu_b= \sqrt{m_B^2-m_b^2}\simeq 2.4$ GeV, $M'^2=10 \pm 2$ GeV$^2$,  
$f_{3\pi}(\mu_b)=0.0026$ GeV$^2$ \cite{BF1,Zhitnitsky:1985} and 
$\delta^2(\mu_b)=0.17$ GeV$^2$ \cite{SVZN,Chernyak}.
The shapes of distribution amplitudes are varied between  
the BF ansatz of nonasymptotic coefficients \cite{BF1}  
(where, in particular, $a_2(1 \mbox{GeV})=0.44$, $a_4(1 \mbox{GeV})=0.25 $ and all $a_{n> 4}=0$) and purely asymptotic form. 
With this input one, in particular, 
obtains \cite{KRWWY00}
\be
f^+_{B\pi}(0)=0.28 \pm 0.05.
\ee 
To quantify the magnitude of the nonfactorizable  soft-gluon 
effect, we introduce the ratio: 
\be
\frac{\lambda_{E}(\bar{B}^0_d\to\pi^+\pi^-)}{m_B}
\equiv \frac{A^{(\widetilde{O}_1)}_{E}(
\bar{B}^0_d\to\pi^+\pi^-)}{A^{(O_1)}_{E}( \bar{B}^0_d\to\pi^+\pi^-)},
\label{lambdaE}
\ee
and obtain the following estimate
\be 
\lambda_{E}(\bar{B}^0_d\to\pi^+\pi^-)= 0.05 \div 0.15 ~\mbox{GeV}\,,
\label{lambdaEnum}
\ee
The quoted interval corresponds to the variation of all input parameters
within adopted ranges and adding linearly the corresponding 
uncertainties. The parameter $\lambda_{E}$ is 
independent of $f_B$ and  
less sensitive to the input than the individual matrix elements
in Eq.~(\ref{lambdaE}). The main uncertainty in $\lambda_E$
is introduced by the nonperturbative parameters $f_{3\pi}$ and $\delta^2$. 
They are extracted from two-point sum rules with a limited accuracy
of about $\pm 30\%$. 

The soft-gluon contribution to 
the matrix element $\langle\pi^+\pi^-| \widetilde{O}_1 | \bar{B}^0_d \rangle$ 
turns out to be very small, at the 
level of $1\div 2\%$ of the factorizable matrix element 
$\langle\pi^+\pi^- | O_1 | \bar{B}^0_d \rangle$. In fact, this effect 
is at the same level as the twist 3,4 contributions of 
quark-antiquark-gluon distribution amplitudes to the LCSR for the 
form factor $f_{B\pi}^+$ \cite{BKR}.

\section{ Soft-gluon correction to the QCD factorization formula}

For the $\bar{B}^0_d\to \pi^+\pi^-$ amplitude, 
the QCD factorization 
formula reads \cite{BBNS99}:
\bea
{\cal A}(\bar{B}^0_d\to \pi^+\pi^-)&\equiv& 
\langle \pi^-\pi^+|H_W |\bar{B}^0_d \rangle
=i\frac{G_F}{\sqrt{2}}V_{ub}V^*_{ud}f_\pi f_{B\pi}^+(0)m_B^2
\nonumber
\\
&&\times\left\{c_1(\mu) +\frac{c_2(\mu)}{3}
+\frac{\alpha_s}{9\pi} c_2(\mu)F(\mu) + 
O \left( \frac{\Lambda_{QCD}}{m_b} \right) \right\}\,, 
\label{qcdfact}
\eea
where, in order to simplify the discussion, we neglect 
the matrix elements of penguin operators/topologies. 
The hard nonfactorizable contributions 
of $O(\alpha_s)$ parametrized by $F$  
have been explicitly calculated \cite{BBNS99},
in a form of the convolution of hard-scattering kernels with the 
pion and $B$-meson distribution amplitudes \footnote{
For the $B\to D^{(*)}\pi$ a similar calculation has been done  
in Ref.~\cite{PW88}.}.   

The  form factor $f^+_{B\pi}$ in Eq.~(\ref{qcdfact}) includes
not only hard, but also soft contributions \footnote{
This important feature distinguishes the QCD factorization approach 
from other analyses of $B\to \pi\pi$ in perturbative QCD \cite{Brodsky,Li}.}.
The latter are not accessible in QCD perturbation theory. 
Therefore, $f^+_{B\pi}$ represents an external input to the 
QCD factorization formula.
Currently, this form factor is calculated in  
lattice QCD using  certain extrapolations. It can also be obtained from LCSR, 
as discussed above. The results of both approaches generally agree 
with each other (see Ref.~\cite{Bec} for a recent comparison).  
The pion distribution amplitudes, at least 
their nonasymptotic terms, e.g., the
coefficients $a_n$ in Eq.~(\ref{gegenb}), represent another set of inputs.
They, in principle, can be determined from lattice QCD \cite{Agliettietal}.
Another possibility to estimate and/or constrain 
the nonasymptotic parts of the pion distribution amplitudes 
is provided by fitting the LCSR prediction on the pion electromagnetic form 
factor \cite{BKM} to experimental data. 
Yet another external input, used to calculate the 
hard spectator contribution to $F$ in Eq.~(\ref{qcdfact}),
is the $B$-meson distribution amplitude, which is currently taken from 
models. 

The last term in Eq.~(\ref{qcdfact}) symbolizes 
contributions of soft nonfactorizable effects. Their magnitude
cannot be calculated within the QCD factorization approach. 
The $1/m_b$ suppression of these nonperturbative effects 
is derived \cite{BBNS99} from the infrared behavior 
of perturbative diagrams. Therefore, in order to control the 
accuracy of QCD factorization  at finite $m_b$, one has 
to estimate soft nonfactorizable corrections in a certain 
nonperturbative framework and confirm their $1/m_b$ suppression.

The LCSR approach described in the previous sections 
yields for the $\bar{B}^0_d\to \pi^+\pi^-$ amplitude an expression 
very similar to Eq.~(\ref{qcdfact}):
\bea
{\cal A}(\bar{B}^0_d\to \pi^+\pi^-)_{LCSR} = 
i\frac{G_F}{\sqrt{2}}V_{ub}V^*_{ud}f_\pi [f_{B\pi}^+(0)]_{LCSR}~m_B^2 
\Bigg\{c_1(\mu) +\frac{c_2(\mu)}{3}
\nonumber
\\
+
2c_2(\mu) \left( \frac{\lambda_{E}(\bar{B}^0_d\to\pi^+\pi^-)}{m_B}
+\!\!\!\sum_{i=A,P,PA}\frac{
\lambda_{i}(\bar{B}^0_d\to\pi^+\pi^-)}{m_B}+O(\alpha_s) \right) \Bigg\}\,, 
\label{ampl}
\eea
In the nonfactorizable part, 
in addition to the soft correction in the emission topology 
calculated in Sect. 3, we indicate the contributions 
to the hadronic matrix element of $\widetilde{O}_1$ which still have to be 
calculated. In particular, the terms proportional to 
$\lambda_{A,P,PA}$ denote soft nonfactorizable corrections due 
to annihilation, penguin, and penguin annihilation topologies. 
We tacitly assume that all these effects are at least of $O(1/m_b)$. 
That has to be confirmed by a direct calculation. Finally, 
$O(\alpha_s)$ in Eq.~(\ref{ampl})
indicates the contribution of hard nonfactorizable 
effects originating from the diagrams in Fig.2 and from analogous diagrams 
of other topologies. All these terms can be calculated one by one from the 
underlying correlation function (\ref{corr}) as it was done 
for $\lambda_E$ . 
Moreover, the calculation is performed using the same framework and input 
as for the form factor $[f_{B\pi}^+(0)]_{LCSR}$.  
We stress that in LCSR there is no need 
to introduce the $B$-meson distribution amplitude because 
the heavy meson is interpolated  by a current in the correlation function.
In particular, the hard-gluon exchange with the spectator quark corresponds 
to the diagrams in Fig.~2c, as already explained in Sect. 3.2.

The completion of Eq.~(\ref{ampl}) is a difficult calculational task. 
Before it is fulfilled, one can use the evaluated 
$\sim\lambda_E/m_B$ term interpreting it as a soft 
correction to the  QCD factorization formula, 
that is, replacing $O(\Lambda_{QCD}/m_b) \to 2c_2\lambda_E/m_B$
in Eq.~(\ref{qcdfact}).  
To compare the soft and hard nonfactorizable corrections, 
we take our estimate (\ref{lambdaEnum}) and 
calculate the short-distance part 
of Eq.~(\ref{qcdfact}) numerically. For consistency, 
we adopt $\mu=\mu_b$, adjusting the QCD factorization formula to 
the characteristic scale 
at which the hadronic matrix elements have been calculated from LCSR. 
Furthermore, we use $c_{1,2}(\mu_b)$ 
in NLO and in the NDR scheme,  
$\alpha_s(\mu_b)=0.279$ corresponding to $\alpha_s(m_Z)= 0.118 $,
and the expression for $F$  given in Ref.~\cite{BBNS99}. In the latter,
we adopt for the integral $\int_0^1 \Phi_B(\xi)/\xi=m_B/\lambda_B$ 
over the $B$ meson distribution amplitude $\Phi_B(\xi)$  
the same parameter $\lambda_B=0.3$ GeV as in Ref.~\cite{BBNS99}.
The following estimates are obtained for 
the separate terms in Eq.~(\ref{qcdfact}): 
\bea \left[ c_1(\mu_b)+\frac{c_2(\mu_b)}3 \right]+
\left[ \frac{\alpha_s}{9\pi} c_2(\mu_b)F(\mu_b)\right]+
\left[ 2c_2(\mu_b)
\frac{\lambda_{E}(\bar{B}^0_d\to\pi^+\pi^-)}{m_B}\right]
\nonumber
\\
= 
[1.03] +[(-0.007 \div +0.01)+0.03i]-[0.005 \div 0.015 ]\,,
\label{a1}
\eea
where the interval 
for the real part in the second bracket on r.h.s.  
corresponds to varying  $\varphi_\pi(u,\mu_b)$ 
between the BF and asymptotic form in the expression for $F$. 
\footnote{We parenthetically notice that 
the real part of the coefficient $F$ in Eq.~(\ref{qcdfact})
is quite sensitive to the shape of $\varphi_\pi(u)$, whereas 
the imaginary part 
is independent of this shape, being simply proportional to the normalization 
of the distribution amplitude.} This interval 
has only a small correlation with the range in the third 
bracket on r.h.s. of Eq.~(\ref{a1}). The latter corresponds to
the variation of $\lambda_E$ in Eq.~(\ref{lambdaEnum})
and is to a large extent caused by the uncertainty 
in the nonperturbative parameters $f_{3\pi}$ and $\delta^2$.

We find that the hard and soft nonfactorizable corrections are 
of the same order and their overall effect varies within $\pm$2\% 
of the factorizable amplitude. 
Thus, in the $\bar{B}^0_d\to \pi^+\pi^-$ channel both effects are 
very small and do not influence the decay amplitude, which is predominantly 
factorizable.
The situation is different in the case of the $c_2+c_1/3$ combination 
relevant for the colour-suppressed $B\to\pi^0\pi^0$ decay: 
\bea \left[ c_2(\mu_b)+\frac{c_1(\mu_b)}3 \right]+
\left[ \frac{\alpha_s}{9\pi} c_1(\mu_b)F(\mu_b)\right]+
\left[ 2c_1(\mu_b)
\frac{\lambda_{E}(\bar{B}^0_d\to\pi^+\pi^-)}{m_B}\right]
\nonumber
\\
= [0.103] +[(0.03\div -0.04)-0.104i]+[0.021 \div 0.064]\,.
\label{a2}
\eea
Here the nonfactorizable corrections have a noticeable impact 
on the amplitude, both on its magnitude and phase.

We conclude that soft nonfactorizable 
effects are as important as the $O(\alpha_s)$ 
nonfactorizable corrections in the QCD factorization formula.
To be cautious, the numbers presented in Eqs.~(\ref{a1}), (\ref{a2})
cannot yet be considered as final predictions to be 
used in the phenomenological analysis. 
Adding to $\lambda_E$ the remaining soft effects, due to annihilation
and penguins, will most probably cause further changes in  
the decay amplitudes.
One also has to stress that the soft and hard effects 
in Eqs.~(\ref{a1}) and (\ref{a2})
are calculated in different approaches. It is more safe and consistent 
to estimate all nonfactorizable effects using one and the same method, 
and the LCSR approach offers such a possibility. 
 
\section{Earlier applications of QCD sum rules} 

There have been several attempts to calculate nonleptonic decays
of heavy mesons using OPE and QCD sum rules
\cite{BS87,BS93,nonfactSR} and focusing on nonfactorizable 
matrix elements. An  important prediction of this approach, 
as emphasized in 
Ref.~\cite{KRHF}, is the channel-dependence of nonfactorizable effects.
However, certain complications manifested themselves 
in the derivation procedure of 
the sum rules.  The original method \cite{BS87} of four-point 
correlation functions, applied to two-body $D$ decays, 
was plagued by a  presence of light-hadron states  
in the dispersion relation for the heavy-meson channel. These states, 
as explained in Sect. 2, originate due to the absence of an external 
momentum in the $b$-quark decay vertex. 
A different
procedure suggested in Ref.~\cite{BS93} for $B\to D \pi$ (and used in 
Ref.~\cite{nonfactSR} for other channels) avoids this problem. 
If we apply the latter procedure to $B\to \pi\pi$, the starting point 
is then a two-point hadronic matrix element 
$M_{B\pi}\equiv i\int d^4x\,\exp(ipx)\langle \pi |T\{ j^{(\pi)}_{\alpha 5}(x)O(0) \}|B\rangle$. 
The expansion of the operator product in a series of local operators,
\be  
T\{j^{(\pi)}_{\alpha 5}(x)O(0)\} = \sum_k C_{\alpha k}(x)O^{eff}_k(0),
\label{expan}
\ee
reduces $M_{B\pi}$ to a set of simpler matrix elements, 
$\langle \pi | O^{eff}_k |B\rangle $, which have to be calculated 
separately using, e.g., QCD sum rules. Only the lowest-dimension 
operator is usually taken into account for $O=O_{1,2}$. 
Matching the result of this calculation with the 
dispersion relation for $M_{B\pi}$ 
in the channel of the current $j^{(\pi)}_{\alpha 5}$, one finally
obtains the two-pion matrix element $\langle \pi \pi | O |B\rangle$.
This estimate is rather ambiguous, because   
quark-hadron duality cannot be applied to the 
hadronic matrix elements involved in this procedure. 
The method suggested in this paper avoids 
any unwanted contamination of the dispersion relation, due to the 
auxiliary external four-momentum in the $b$ -decay vertex. 
With this additional element, the sum rule for the 
hadronic matrix element is directly obtained from the initial 
three-point correlation function, as demonstrated in Sect.~2. One 
therefore does not need to follow the two-stage method of 
effective operators.

Apart from these essential improvements, 
there is also a very important difference between LCSR derived in 
this paper and sum rules obtained before. At all stages of the 
calculational procedure described in Sect.~2,3 we use 
the light-cone OPE. In contrast,
the sum rules obtained from four-point correlation functions \cite{BS87} 
are based on the short-distance expansion in local condensates.
Similarly, the method of effective operators \cite{BS93} 
employs a truncated short-distance expansion (\ref{expan}). 
This might be a reasonable approximation  for a specific kinematics 
of $B\to D \pi$ considered in Ref.~\cite{BS93}, where the pion energy 
does not scale with the heavy mass (see also a discussion on this point 
in the second paper of Ref.~\cite{BBNS99}). However, as already emphasized 
in Ref.~\cite{BS93}, if the non-spectator hadron carries a large energy 
$\sim m_b$, as in $B\to \pi\pi$, the resummation of local operators with the same twist 
is necessary.  In fact, such a resummation is implicitly 
performed within the procedure
of the light-cone OPE employed in Sect.~3. 

It is instructive to demonstrate what happens 
if one employs a truncated short-distance expansion, instead of the 
light-cone one, 
in the correlation function (\ref{corr}). We
consider again the diagram in Fig. 1b with the effective operator 
$\widetilde{O}_1$, and perform a short-distance expansion 
similar to the one in Eq.~(\ref{expan}): 
\be  
i\int d^4y~e^{i(p-k)y}T\{j^{(\pi)}_{\alpha 5}(y)\widetilde{O}_1(0)\} = 
C_{\alpha\mu\lambda\rho}(p-k)\bar{u}(0)\Gamma^\mu\widetilde{G}^{\lambda\rho}(0)b(0)\,,
\label{expan2}
\ee
retaining only one local operator with the lowest dimension.
To obtain the corresponding Wilson coefficient $C_{\alpha \mu\lambda\rho}$ 
one simply has to put $v\to 0$ in the argument of the gluon field in the quark 
propagator (\ref{prop}). This local limit  corresponds to the quark propagating
in the static gluon field, a usual approximation for the condensates 
in the conventional QCD sum rules \cite{SVZ}.
Thus, in Eq.~(\ref{expan2})
an infinite series of quark-gluon operators of the 
same twist containing derivatives of the gluon field is neglected.  
Physically, this approximation, corresponding to a vanishing four-momentum 
of the soft gluon, cannot be justified, because in the diagram of Fig. 1b 
the momentum of the on-shell gluon is 
in the range $0<v\alpha_3q <q $, where $q$ is of the order of 
the virtual quark momenta. 

After substituting the expansion (\ref{expan2}) back in the correlation
function (\ref{corr}), we proceed
using the light-cone expansion of the product $O_{eff}(0)j_5^{(B)}(x)$
and confining ourselves by the twist 3 contribution.
The result is given 
by the same expression as in Eq.~(\ref{tw3F}),
but with $v=0$ in the denominator:
\bea
\left(F^{(\widetilde{O}_1)}_{E,tw3}\right)_{\mbox{short dist.}}
=\frac{m_b f_{3\pi}}{4\pi^2(-(p-k)^2)}
\int {\cal D}\alpha_i ~
\frac{\varphi_{3\pi}(\alpha_i)}{(m_b^2-(p-q)^2(1-\alpha_1))}
\nonumber
\\
\times \left[\frac32(q\cdot k) +q\cdot(p-k)\right] (q\cdot(p-k))\,.
\label{tw3FSD}
\eea
We emphasize that in this modified procedure, the short-distance 
expansion has only been partially used. Nevertheless, there
are  important changes
in the resulting expression (\ref{tw3FSD}) as compared with Eq.~(\ref{tw3F}).
The dependence of $F_{\alpha E}^{(\widetilde O_1)}$ on the variable $P^2$ 
drops out, and the dependence on $(p-k)^2$ reduces to 
a simple factor $1/(p-k)^2$, with 
an imaginary part proportional to $\delta(s-(p-k)^2)$. 
It is therefore difficult to attribute 
certain duality counterparts to the contributions of the pion and 
hadronic continuum, while  matching the short-distance version of
$F_{\alpha E}^{(\widetilde O_1)}$ with the dispersion relation 
(\ref{disp}).  Let us, nevertheless, continue 
the derivation of the sum rule using Eq.~(\ref{tw3FSD})
and assuming a complete pion dominance in the dispersion relation. 
Following the same procedure as described in Sect.~2,
we obtain the hadronic matrix element
\bea
&&A^{(\widetilde{O}_1)}_{E}( \bar{B}_d^0\to\pi^+\pi^-)_{\mbox{short dist.}}
= -\frac{if_{3\pi}}{16\pi^2f_\pi}\left(\frac{m_b^3}{2f_B}\right)
\int\limits_{u_0^B}^1 
\frac{du}{u^2}e^{\,m_B^2/M'^2-m_b^2/uM'^2}
\nonumber
\\
&&\times\int\limits_0^u dv~\varphi_{3\pi}(1-u,u-v,v)
\left( 3-\frac{m_B^2}{m_b^2}u \right)\,,
\label{Asd}
\eea
where only the twist 3 part is taken into account. 
We have checked that a very similar result is 
obtained employing the two-step procedure
\cite{BS93} based on the effective operators, if one uses 
the approximation (\ref{expan2}) for the expansion 
(\ref{expan}) and calculates the $B\to \pi$ matrix element
of the effective operator separately, using the LCSR.

In the heavy quark limit, evaluated using Eq.~(\ref{hqet}),
the r.h.s. of Eq.~(\ref{Asd}) is proportional to $ m^{3/2}$, enhanced 
by one power of $m_b$ in comparison with the twist-2 factorizable 
matrix element (\ref{Afacttw2}). Note that the latter 
is obtained from the diagram in Fig.~1a containing free-quark 
propagators, and there is no difference between 
short-distance and light-cone expansions in this case. The  situation 
closely resembles QCD sum rules for heavy-to-light 
form factors based on the local condensate expansion. In these sum rules, 
the momenta of  soft quarks and gluons  forming vacuum condensates are 
also neglected yielding an enhancement of condensate terms 
at $m_b\to \infty$ as compared with the leading-order perturbative terms
(for more details, see the discussion in Ref.~\cite{BallBraun}). 
Numerically, the nonfactorizable correction $\lambda_E/m_B$ 
calculated from Eq.~(\ref{Asd}) is almost an order of magnitude larger
than the estimate (\ref{lambdaEnum}) and has a different sign. 
Note that the sign difference in  the nonfactorizable 
matrix element obtained using the new LCSR procedure 
is a consequence of the $P^2$ dependence in the correlation function 
and emerges after the analytic continuation $(P^2<0) \to m_B^2$. 

We conclude that even a limited use of the short-distance expansion 
in the correlation function (\ref{corr})
yields a sum rule, where it is difficult to separate the ground-state 
contribution applying the usual duality 
approximation and where the higher-twist terms are enhanced 
at $m_b\to \infty$ with respect to the leading twist-2 term.

\section{ Conclusions}

In this paper, a new method has been suggested to calculate the
hadronic matrix elements relevant for two-body nonleptonic $B$ decays.
The approach is based on the light-cone expansion of a three-point
corelation function and on the analytic 
continuation from the spacelike to the timelike region. 
The resulting hadronic matrix element 
contains a factorizable part as well as soft and hard
nonfactorizable corrections. All these contributions 
can be calculated within one framework using a common input. 

The nonfactorizable soft-gluon exchange correction 
calculated for the $B\to \pi\pi$ channel is our main result. 
Importantly, the $1/m_b$ suppression of this effect has been 
explicitly reproduced,
in agreement with the expectation of the QCD factorization
approach. Moreover, it is encouraging that the soft correction
in $\bar{B}^0\to \pi^+\pi^-$ turns out to be numerically small, 
suggesting that there are no unexpectedly large power corrections
to QCD factorization in this channel. 

The accuracy of the LCSR method is limited, as usual in QCD sum rules, 
since one has to rely on the quark-hadron duality,
both in the pion and $B$ channels. Nevertheless, one may argue 
that the duality approximation works reasonably well having in mind, 
e.g. an agreement of  
the LCSR predictions for the $B\to \pi$ form factors 
with the lattice QCD results. 
It is also important that uncertainties  of the sum rule 
predictions can be assessed by varying the input parameters within the 
currently adopted limits. These parameters are  universal, therefore,
their accuracy can be improved in future by direct determinations 
on the lattice  or by fitting QCD sum rules for experimentally measured 
hadronic characteristics to the data.

The same method, with minor modifications, can be used for 
all other channels 
where $B$ meson decays either into two light hadrons, or into one 
charmed hadron (or charmonium state) and  
a light hadron. The important examples in the latter case 
are $B\to D^{(*)} \pi, J/\psi K^{(*)}$. The only limitation is  
that at least one of the light-quark 
hadrons in the final state has to be a pseudoscalar or vector meson, 
for which the  light-cone distribution amplitudes are better known.     

\bigskip

{\bf Acknowledgments}\\

I am grateful to P.~Ball, J.~Bijnens, V.~Braun,
G.~Buchalla, P.~Colangelo, T.~Mannel, 
R.~R\"uckl, Yu.~Shabelski and  M.~Shifman
for useful discussions and comments. 
This paper was completed at the Lund University,
where my work is supported by The Swedish Foundation
for International Cooperation in Research and Higher Education (STINT).

\end{document}